\RequirePackage{snapshot}
\documentclass[10pt,conference]{IEEEtran}
\usepackage[utf8]{inputenc}

\usepackage[cmex10]{amsmath}
\usepackage{amsfonts}
\usepackage{ifdraft}
\usepackage{datetime}

\usepackage[caption=false,font=footnotesize]{subfig}

\newtheorem{theorem}{Theorem}
\newtheorem{lemma}[theorem]{Lemma}

\newcommand\goesto{\rightarrow}
\newcommand\ra{\rightarrow}

\newcommand{\Sh}{{\hat{S}}}
\newcommand{\Qh}{{\hat{Q}}}
\newcommand{\Eh}{{\hat{E}}}

\newcommand{\R}{\mathbb{R}}
\newcommand{\Z}{\mathbb{Z}}
\newcommand{\E}{\mathbb{E}}
\newcommand{\N}{\mathcal{N}}

\newcommand{\I}{\mathcal{I}}

\newcommand{\cE}{\mathcal{E}}
\newcommand{\Err}{{\cE}}

\newcommand{\vect}[1]{\mathbf{#1}}

\newcommand{\X}{\vect{X}}

\newcommand{\ssq}{{\sigma_S^2}}
\newcommand{\sz}{{\sigma_Z}}
\newcommand{\szq}{{\sz^2}}
\newcommand{\se}{{\sigma_E}}
\newcommand{\seq}{{\se^2}}

\newcommand\eg{\emph{e.g.}}

\newcommand\e{\epsilon}

\newcommand\snr{\textsc{snr}}

\newcommand{\Eqi}{{\Err_{Q,i}}}
\newcommand{\Ee}{{\Err_E}}

\newcommand{\pmin}{{p_{\min}}}
\newcommand{\mse}{{\E[(\Sh - S)^2]}}
\newcommand{\msecond}{{\E[(\Sh - S)^2 | s]}}
\newcommand{\msecondd}{{\E[(\Sh - S - \Delta)^2 | s + \Delta]}}
\newcommand{\dhsq}{{\left(\frac{\Delta}{2}\right)^2}}
\newcommand{\intabd}{{\int_A^{B-\Delta}}}

\newcommand{\pe}{P_e}

\newcommand\deq{\stackrel{\mathrm{def}}{=}}

\DeclareMathOperator\Int{int}
\DeclareMathOperator\var{Var}

\begin{document}

\title{A Tight Bound on the Performance of a Minimal-Delay Joint Source-Channel
Coding Scheme\looseness=-1}
\author{\IEEEauthorblockN{Marius Kleiner, Bixio Rimoldi}
\IEEEauthorblockA{School of Computer and Communication Sciences\\
Ecole Polytechnique F\'ed\'erale de Lausanne\\
CH-1015 Lausanne, Switzerland\\
E-mail: firstname.lastname@epfl.ch}}

{\usdate \maketitle}

\ifdraft{
\begin{center}
\huge  \textbf{\emph{Compiled {\mdyyyydate\today} \xxivtime}}
\end{center}
}

\begin{abstract}
  An analog source is to be transmitted across a Gaussian channel in
  more than one channel use per source symbol. This paper derives a lower bound
  on the asymptotic mean squared error for a strategy that consists of
  repeatedly quantizing the source, transmitting the quantizer outputs in the
  first channel uses, and sending the remaining quantization error uncoded in
  the last channel use. The bound coincides with the performance achieved by a
  suboptimal decoder studied by the authors in a previous paper, thereby
  establishing that the bound is tight. 
\end{abstract}

\section{Introduction}

This paper gives performance limits of a certain class of encoders for the
transmission of a discrete-time memoryless analog source across a
discrete-time memoryless Gaussian channel, where the channel can be used $n$
times for each source symbol. The parameter~$n$ is arbitrary but fixed, given as
part of the problem statement.

It is well known that if  the channel noise
has variance~$\szq$ then the average transmit power~$P$ and the average
mean-squared error $D$ of any communication scheme for this scenario are related
by
\begin{equation}
  \label{eq:shannonlimit}
  R(D) \le nC(P),
\end{equation}
where $R(D)$ is the rate-distortion function of the source under squared-error
distortion and $C(P) = 0.5 \log( 1 + P/\szq)$ is the cost-constrained capacity
of the channel (see \eg~\cite{CoverT1991}). If the source has finite
differential entropy~$h(S)$ then the rate-distortion function satisfies
\begin{equation*}
  R(D) \ge h(S) - 0.5 \log(2\pi e D).
\end{equation*}
Applying this bound to~\eqref{eq:shannonlimit} and inserting the capacity
formula yields
\begin{equation}
  \label{eq:Dlowerbound}
  D \ge \frac{2^{2h(S)}}{2\pi e} (1 + P/\szq)^{-n}
\end{equation}
or $D \ge c(1 + \snr)^{-n}$, where we have defined $\snr = P/\szq$ and $c =
2^{2h(S)} / 2\pi e$. As $\snr \ra \infty$, the squared error distortion scales
thus at best as $\snr^{-n}$. 

In this paper we study a communication scheme for this scenario that is
extremely simple to implement and has minimal delay, in the sense that it
encodes and transmits a single source symbol at a time. It works by quantizing
the source and then repeatedly quantizing the quantization error; the quantized
points are sent across the first $n-1$ channel uses and the last quantization
error is sent uncoded in the $n^{\text{th}}$ channel use. 

We show that no matter how the quantization resolution is chosen (as a function
of the SNR) and regardless of the decoder used, the mean squared error achieved
by this scheme cannot decay faster than $\snr^{-n}(\log\snr)^{n-1}$.  This
asymptotic lower bound coincides with the achievable performance of a suboptimal
decoder that we have studied in a previous paper~\cite{KleinerR2009}; it is
therefore tight.

%

\medskip
Transmission schemes of the kind proposed here have been considered
before. Indeed, one of the first schemes to transmit an analog source across two
uses of a Gaussian channel was suggested by Shannon~\cite{Shannon1949}.
Generalizing Shannon's ideas, Wozencraft and Jacobs~\cite{WozencraftJ1965}
provided the foundations to analyze source-channel mappings as curves in
$n$-dimensional space.  Ziv~\cite{Ziv1970} found important theoretical
limitations of such mappings.  

Much of the later work is due to Ramstad and his coauthors (see
\eg~\cite{Ramstad2002}, \cite{CowardR2000a}, \cite{HeklandFR2009}). A proof that
the performance of minimal-delay codes is strictly smaller than that of codes
with unrestricted delay when $n>1$ was given in 2008 by Ingber et
al.~\cite{IngberLZF2008}.

For $n=2$, the presented scheme is almost identical to the HSQLC scheme by
Coward~\cite{Coward2001}, which uses a numerically optimized quantizer,
transmitter and receiver to minimize the mean-squared error (MSE) for finite
values of the SNR. Coward conjectured that the right strategy for $n > 2$ would
be to repeatedly quantize the quantization error from the previous step, which
is exactly what we do here.

Another closely related communication scheme is the \emph{shift-map} scheme due
to Chen and Wornell~\cite{ChenW1998}.  Vaishampayan and
Costa~\cite{VaishampayanC2003} showed in their analysis that it achieves a
squared error that scales as $\snr^{-n+\e}$ for any fixed $\e > 0$ if the
relevant parameters are chosen correctly as a function of the SNR. Up to
rotation and a different constellation shaping, the shift-map scheme is in fact
virtually identical to the one used here, a fact that was pointed out recently
by Taherzadeh and Khandani~\cite{TaherzadehK2008}. This suggests that its
performance is also limited by the bound derived here.

%
Other hybrid schemes, such as the one by Shamai et al.~\cite{ShamaiVZ1998} or
that of Mittal and Phamdo~\cite{MittalP2002}, use long block codes for the
digital phase and are therefore not directly comparable with minimum delay
schemes.

\medskip
The rest of this paper is organized as follows. Section~\ref{sec:commscheme}
describes our transmission strategy, in Section~\ref{sec:achievable} we quote
the achievability result from our previous paper, and finally
Section~\ref{sec:distlowerbounds} contains our derivation of the mean squared
error lower bound.

\section{Proposed Communication Scheme}
\label{sec:commscheme}

To encode a single source letter $S$ into $n$~channel input symbols $X_1$,
\dots, $X_n$ we proceed as follows. Define $E_0 = S$ and recursively compute
the pairs $(Q_i, E_i)$ as
\begin{align}
  Q_i &= \frac{1}{\beta} \Int(\beta E_{i-1}) \nonumber \\
  E_i &= \beta (E_{i-1} - Q_i) \label{eq:QEdef}
\end{align}
for $i = 1$, \dots, $n-1$ where $\Int(x)$ is the unique integer~$i$ such that
\begin{equation*}
  x \in \left[ i - \frac12 , i + \frac12 \right)
\end{equation*}
and $\beta$ is a scaling factor that \emph{grows with the SNR}
in a way to be determined later. $Q_i$ is thus a quantized version of the
previous round's quantization error, and $E_i$ is the new quantization
error scaled up to lie in $[-1/2, 1/2)$.
%
Note that the map $S \mapsto (Q_1, \dots, Q_{n-1}, E_{n-1})$ is one-to-one with
the inverse given by
\begin{equation}
  \label{eq:unwraprec}
  S = \sum_{i=1}^{n-1} \frac{1}{\beta^{i-1}} Q_i + \frac{1}{\beta^{n-1}}
  E_{n-1}.
\end{equation}

We determine the channel input symbols $X_i$ from the $Q_i$ and from $E_{n-1}$
according to 
\begin{align}
  X_i &= \sqrt{\frac{P}{\ssq + \delta}} Q_i \quad \text{for $i = 1$,
  \dots, $n-1$
  and} \nonumber \\
  \label{eq:Xdef}
  X_n &= \sqrt{\frac{P}{\seq}} E_{n-1},
\end{align}
where $\seq = \var(E_{n-1})$, and where $\delta > 0$ is some small number. As
shown in~\cite{KleinerR2009}, this ensures that $\E[X_i^2] \le P$ for all~$i$
and for $\beta > \beta_0$ (where $\beta_0$ depends on $\delta$).  Since we are
interested in the large SNR regime and since we have defined $\beta$ to grow
with the SNR, we can assume for the remainder that the power constraint is
satisfied.

\section{Achievable Performance}
\label{sec:achievable}

In this section we quote the relevant results of our earlier
paper~\cite{KleinerR2009}, which imply that a suboptimal decoder achieves a mean
squared error that scales at least as $\snr^{-n}(\log\snr)^{n-1}$. While we only
considered Gaussian sources in that paper, we actually never used the
distribution of the source when we derived the bounds there, so they hold for
general continuous sources of bounded variance.

\subsection{Suboptimal Decoder}

The encoder outputs $X_i$ are transmitted across the channel, producing at the
channel output the symbols
\begin{equation*}
  Y_i = X_i + Z_i, \quad i = 1, \dots, n,
\end{equation*}
where the $Z_i$ are iid Gaussian random variables of variance~$\szq$. 
To estimate $S$ from  $Y_1$, \dots, $Y_n$, the decoder first
computes separate estimates $\Qh_1$, \dots, $\Qh_{n-1}$ and $\Eh_{n-1}$, and
then combines them to obtain the final estimate~$\Sh$. 

To estimate the $Q_i$ we use a maximum likelihood (ML) decoder, which yields the
minimum distance estimate
\begin{equation}
  \label{eq:mldecoder}
  \Qh_i = \frac{1}{\beta} \arg \min_{j\in \Z} \left| \sqrt{\frac{P}{\ssq
  + \delta} } \frac{j}{\beta} - Y_i \right|.
\end{equation}
To estimate $E_{n-1}$, we use a linear minimum mean-square error (LMMSE)
estimator (see \eg~Scharf~\cite[Section~8.3]{Scharf1990}), which computes
\begin{equation}
  \label{eq:lmmse}
  \Eh_{n-1} = \frac{\E[E_{n-1} Y_n]}{\E[Y_n^2]} Y_n.
\end{equation}
Finally we use~\eqref{eq:unwraprec} to obtain
\begin{equation}
  \label{eq:unwrapestim}
  \Sh = \sum_{i=1}^{n-1} \frac{1}{\beta^{i-1}} \Qh_i + \frac{1}{\beta^{n-1}}
  \Eh_{n-1}.
\end{equation}

\subsection{Upper Bounds on the Mean Squared Error}

Using the suboptimal decoder described in the previous section, $\E[(S-\Sh)^2]$
can be broken up into contributions due to the errors in decoding $Q_i$ and
$E_{n-1}$ as follows. From~\eqref{eq:unwraprec} and~\eqref{eq:unwrapestim}, the
difference between $S$ and $\Sh$ is
\begin{equation*}
  S - \Sh = \sum_{i=1}^{n-1} \frac1{\beta^{i-1}} (Q_i - \Qh_i) + \frac1{\beta^{n-1}}
  (E_{n-1} - \Eh_{n-1}).
\end{equation*}
The error terms $Q_i - \Qh_i$ depend only on the noise of the respective channel
uses and are therefore independent of each other and of $E_{n-1} - \Eh_{n-1}$,
so we can write the error variance componentwise as
\begin{equation}
  \label{eq:totalerror}
  \E[(S-\Sh)^2] = \sum_{i=1}^{n-1} \frac{1}{\beta^{2(i-1)}} \Eqi +
  \frac{1}{\beta^{2(n-1)}} \Ee, 
\end{equation}
where $\Eqi \deq \E[(Q_i - \Qh_i)^2]$ and $\Ee \deq \E[(E_{n-1} -
\Eh_{n-1})^2]$.

The following two Lemmata, taken directly from~\cite{KleinerR2009}, give upper
bounds on the two types of errors.  (The $O$-notation is defined in
Appendix~\ref{app:Onotation}.)
\begin{lemma}
  \label{lem:eqbound}
  For each $i = 1$, \dots, $n-1$, 
  \begin{equation}
    \label{eq:eqidecay}
    \Eqi \in O\left(\exp\{-k \snr/\beta^2\}\right),
  \end{equation}
  where $\snr = P/\szq$ and $k > 0$~does not depend on~$\snr$.
\end{lemma}

\begin{lemma}
  \label{lem:eedecay}
  The estimation error of $E_{n-1}$ satisfies
  \begin{equation}
    \label{eq:eedecay}
    \Ee/\beta^{2(n-1)} \in O(\snr^{-1}\beta^{-2(n-1)}). 
  \end{equation}
\end{lemma}

From Lemma~\ref{lem:eqbound}, $\beta^2$ should scale less than linearly
in~$\snr$, otherwise the upper bound would be constant. We therefore let
$\beta^2 = \snr^{1-\e}$, where $\e$ is an arbitrary, strictly positive function
of~$\snr$.  From~\eqref{eq:eqidecay} and~\eqref{eq:eedecay} we have then
\begin{equation*}
  \label{eq:oboundq}
  \Eqi \in O(\exp\{-k \snr^\e\})
\end{equation*}
and
\begin{equation*}
  \label{eq:obounde}
  \Ee/\beta^{2(n-1)} \in O(\snr^{-n+(n-1)\e}).
\end{equation*}
Setting $\e(\snr) = \log((n/k)\log\snr)/\log\snr$ we find that the performance
achieved by the suboptimal decoder satisfies
\begin{equation*}
  \mse \in O(\snr^{-n}(\log\snr)^{n-1}).
\end{equation*}
The next section shows that this is the best achievable scaling for the given
encoder, even if an optimal decoder is used.

\section{Distortion Lower Bounds}
\label{sec:distlowerbounds}

The goal of this section is to lower bound the scaling of the mean squared error
of the transmission strategy described in Section~\ref{sec:commscheme}.

Throughout this section we assume $\beta^2 = \snr^{1-\e}$, where $\e = \e(\snr)$
is a positive function of~$\snr$. This results in no loss of generality, since
for an arbitrary positive function~$f$ we can set $\e(\snr) =
1-\log(f(\snr))/\log\snr$ to get $\beta^2(\snr) = f(\snr)$.

Note that by~\eqref{eq:QEdef}, the $Q_i$ are completely determined by~$S$.
In this section, with a slight abuse of notation, we therefore write $Q_i(s)$ to
denote the value of~$Q_i$ when $S = s$. We use $E_i(s)$ and $X_i(s)$ in a
similar manner. Furthermore, we define $\X(s) = (X_1(s), \dots, X_n(s))$.

The following result, adapted from Ziv~\cite{Ziv1970}, is a key ingredient in
the proofs of the lemmas that follow.

\begin{lemma}
  \label{lem:zivbound}
  Consider a communication system where a con\-tin\-u\-ous-valued source~$S$ is
  encoded into an $n$-dimensional vector $\X(S)$, sent across $n$~independent
  parallel AWGN channels with noise variance~$\szq$, and decoded at the receiver
  to produce an estimate~$\Sh$.  If the density $p_S$ of the source is such that
  there exists an interval $[A,B]$ and a number $p_{\min} > 0$ such that $p_S(s)
  \ge p_{\min}$ whenever $s \in [A,B]$, then for any $\Delta \in [0,B-A)$ the
  mean squared error incurred by the communication system satisfies
  \begin{equation}
    \label{eq:zivbound}
    \E[(\Sh - S)^2] \ge p_{\min} \left(\frac{\Delta}{2} \right)^2 
    \int_A^{B-\Delta} Q(d(s, \Delta) / 2 \sz) ds,
  \end{equation}
  where $d(s, \Delta) \deq \|\vect{X}(s) - \vect{X}(s+\Delta)\|$ and 
  \[Q(x) = \int_x^{\infty} (1/\sqrt{2\pi}) \exp\{-\xi^2/2\} d\xi.\]
\end{lemma}

\begin{IEEEproof}
  See Appendix~\ref{app:zivboundproof}.
\end{IEEEproof}

The next two lemmata provide two different asymptotic lower bounds on the
mean squared error of our transmission strategy, each of which is tighter for a
different class of~$\e$. They hold regardless of the decoder used.  (The
$\Omega$-notation is defined in Appendix~\ref{app:Onotation}.)

\begin{lemma}
  \label{lem:lowerbound1}
  For an arbitrary function $\e(\snr) \ge 0$, the mean squared error satisfies
  \begin{equation*}
    \mse \in \Omega(\snr^{-n + (n-1)\e}).
  \end{equation*}
\end{lemma}

\begin{lemma}
  \label{lem:lowerbound2}
  For an arbitrary function $\e(\snr) \ge 0$, the mean squared error satisfies
  \begin{equation*}
    \mse \in \Omega(\snr^{-1+\e/2} \exp\{-\snr^\e/k\})
  \end{equation*}
  where $k > 0$ does not depend on~$\snr$.
\end{lemma}

\emph{Discussion:} An immediate consequence of the lemmata is that the
theoretically optimal scaling $\snr^{-n}$ is not achievable with the given
encoding strategy: by Lemma~\ref{lem:lowerbound1} this would require $\e = 0$,
but following Lemma~\ref{lem:lowerbound2} the scaling is at best $\snr^{-1}$ if
$\e = 0 $.  More generally, which one of the two lower bounds decays more slowly
and is therefore tighter depends on the scaling of~$\e(\snr)$. How to
choose~$\e(\snr)$ optimally will be the subject of Theorem~\ref{thm:scalinglb}.

\begin{IEEEproof}[Proof of Lemma~\ref{lem:lowerbound1}]
  Assume $\Delta \in [0, \beta^{-(n-1)})$ and define for $j \in \Z$
  \[ \I_j^\Delta = \left[ (j - \frac12 )\beta^{-(n-1)}, 
    (j + \frac12 ) \beta^{-(n-1)} - \Delta \right).\]
  It can be verified from~\eqref{eq:QEdef} that if $s \in \I_j^\Delta$ for
  some~$j$, the following properties hold: 1) $Q_i(s) = Q_i(s+\Delta)$ for
  $i=1$, \dots, $n-1$, and 2) $E_{n-1}(s+\Delta) - E_{n-1}(s) =
  \beta^{n-1}\Delta$.  From~\eqref{eq:Xdef} it follows that
  $s \in \I_j^\Delta$ implies $d(s, \Delta) = \sqrt{P/\seq} \beta^{n-1}\Delta$.

  We now apply Lemma~\ref{lem:zivbound} and restrict the integral to the
  set~$\psi(\Delta) \deq [A,B-\Delta) \cap \bigcup_{j\in\Z} \I_j^\Delta$. The
  lower bound is then relaxed to give
  \begin{equation*}
    \mse \ge \frac{\pmin}{4} \Delta^2 Q(\sqrt{\snr/\seq} \beta^{n-1} \Delta/2)
    \int_{\psi(\Delta)} ds.
  \end{equation*}
  Letting $\Delta = 1/(\sqrt{\snr}\beta^{n-1})$ and $\beta^2 = \snr^{1-\e}$
  yields
  \begin{equation*}
    \mse \ge \frac{\pmin}{4} \snr^{-n+(n-1)\e} Q\left(\frac{1}{2\se}\right)
    \int_{\psi(\Delta)} ds.
  \end{equation*}

  The proof is almost complete, but we still have to show that
  $\int_{\psi(\Delta)}ds$ can be bounded below by a constant for large SNR. The
  length of a single interval~$\I_j^\Delta$ is $\beta^{-(n-1)} - \Delta$. Within
  $[A,B-\Delta)$ there are $(B-A-\Delta)\beta^{n-1}$ such
  intervals. The total length of all intervals~$\I_j^\Delta$ in $[A, B-\Delta)$
  is therefore
  \[ \int_{\psi(\Delta)} ds = (B-A-\Delta)
  (1 - \beta^{n-1}\Delta), \]
  which, for the given values of~$\beta$ and~$\Delta$, 
  converges to $B-A$ for $\snr \ra \infty$ and thus can be lower bounded by a
  constant for $\snr$ greater than some $\snr_0$. With this, the proof is
  complete.
\end{IEEEproof}

\begin{IEEEproof}[Proof of Lemma~\ref{lem:lowerbound2}]
  Observe first that~\eqref{eq:QEdef} implies $Q_1(s + \beta^{-1}) = Q_1(s) +
  \beta^{-1}$ and $E_1(s + \beta^{-1}) = E_1(s)$. Since all $Q_i$ and $E_i$ for
  $i \ge 2$ are by recursion a function of $E_1$ only, $Q_i(s) = Q_i(s +
  \beta^{-1})$ for $i = 2$, \dots, $n-1$, and $E_{n-1}(s) = E_{n-1}(s +
  \beta^{-1})$. Consequently,  $X_i(s) = X_i(s + \beta^{-1})$ for all $i =
  2$, \dots, $n$. By~\eqref{eq:Xdef} and the above, the Euclidean distance
  between $\X(s)$ and~$\X(s+\beta^{-1})$ is therefore
  \begin{equation}
    \label{eq:xbetadist}
    \sqrt{\frac{P}{\ssq + \delta}} |Q_1(s) - Q_1(s+\beta^{-1})| 
    = \sqrt{\frac{P}{\ssq + \delta}} \beta^{-1}.
  \end{equation}

  We now apply Lemma~\ref{lem:zivbound} with $\Delta = \beta^{-1}$. The
  parameter $\beta$ will be chosen to increase with~$\snr$, therefore $\Delta
  \in [0, B-A)$ holds for sufficiently large~$\snr$.

  Using~\eqref{eq:xbetadist}, the resulting bound on the mean squared error is
  \begin{equation*}
    \mse \ge \frac{\pmin}{4} \beta^{-2}
    Q\left (\sqrt{\frac{\snr}{\ssq + \delta}} \frac{\beta^{-1}}{2} \right)
    (B-A-\beta^{-1}
    ).
  \end{equation*}
  Replacing $\beta^2 = \snr^{1-\e}$ and using the fact that $Q(x)$ converges to
  $\exp\{-x^2/2\}/\sqrt{2\pi}x$ for $x \goesto \infty$
  (cf.~\cite{AbramowitzS1964}) we obtain
  \begin{equation*}
    \mse \ge c \snr^{-1 + \e/2} \exp\{-\snr^\e/k\}
  \end{equation*}
  for sufficiently large~$\snr$, with $c$ and $k$ positive constants that do
  not depend on~$\snr$, thus proving the lemma.
\end{IEEEproof}

The following lemma will be used to prove Theorem~\ref{thm:scalinglb}, the main
result of this paper.

\begin{lemma}
  \label{lem:epssolution}
  Define $W(x)$ to be the function that satisfies $W(x)e^{W(x)} = x$ for $x >
  0$.  This function is well defined and is sometimes called the \emph{Lambert
  $W$-function}~\cite{CorlessGHJK1996}. Then for $\snr > 1$ and arbitrary real
  constants $a$, $b>0$, and $k > 0$, 
  \begin{equation}
    \label{eq:epsequation}
    \snr^{a+b\e} = \exp\{-\snr^\e/k\},
  \end{equation}
  if and only if
  \begin{equation}
    \label{eq:epssolution}
    \snr^\e = bk W(\snr^{-a/b} / bk).
  \end{equation}
\end{lemma}

\begin{IEEEproof}
  Let $\snr>1$. Since $\snr^{a+b\e}$ is strictly increasing and
  $\exp\{-\snr^\e/k\}$ is strictly decreasing in~$\e$, there is at most one
  solution to~\eqref{eq:epsequation}.  Assume now $\snr^\e$ is as
  in~\eqref{eq:epssolution}. Then
  \begin{equation*}
    \exp\{-\snr^\e/k\} = \exp\{-b W(\snr^{-a/b}/bk)\}.
  \end{equation*}
  On the other hand,
  \begin{align*}
    \snr^{a+b\e} &= \snr^a \left( bk W(\snr^{-a/b}/bk) \right)^b \\
    &= \left( W(\snr^{-a/b}/bk) / (\snr^{-a/b}/bk) \right)^b.
  \end{align*}
  By definition, $W(x)/x = e^{-W(x)}$, so the above is equal to
  \begin{equation*}
    \snr^{a+b\e} = \exp\{-bW(\snr^{-a/b}/bk)\},
  \end{equation*}
  which proves the claim.
\end{IEEEproof}

The following is the main result of the paper.
\begin{theorem}
  \label{thm:scalinglb}
  For any parameter~$\beta$ and for any decoder, the mean squared error of the
  transmission strategy described in Section~\ref{sec:commscheme} satisfies
  \begin{equation*}
    \mse \in \Omega(\snr^{-n}(\log\snr)^{n-1}).
  \end{equation*}
\end{theorem}

\emph{Discussion:} The asymptotic lower bound on the mean squared error given by
the theorem coincides with the asymptotic performance achieved by the suboptimal
decoder in Section~\ref{sec:achievable}; the bound is therefore asymptotically
tight. 

\begin{IEEEproof}[Proof of Theorem~\ref{thm:scalinglb}]
  Define $l_1(\snr, \e) = \snr^{-n+(n-1)\e}$ and $l_2(\snr,\e) = \snr^{-1+\e/2}
  \exp\{-\snr^\e/k\}$. By Lemmata~\ref{lem:lowerbound1}
  and~\ref{lem:lowerbound2},
  \begin{equation*}
    \mse \in \Omega\big(\max \left( l_1(\snr,\e), l_2(\snr,\e) \right) \big).
  \end{equation*}
  The optimal parameter $\e(\snr)$ is therefore such that for
  any~$\snr$
  \begin{equation}
    \label{eq:epsmax}
    \max\left( l_1(\snr,\e), l_2(\snr,\e) \right)
  \end{equation}
  is minimized. Now for any fixed~$\snr$, $l_1(\snr,\e)$ is increasing in~$\e$,
  and $l_2(\snr,\e)$ is increasing in~$\e$ for $0 \le \e < \xi =
  \log(k/2)/\log\snr$ and decreasing in~$\e$ for $\e \ge \xi$.
  The maximum
  in~\eqref{eq:epsmax} is therefore minimized either for $\e = 0$ or for $\e \ge
  \xi$
  such that $l_1(\e) = l_2(\e)$. As we have remarked earlier, $\e = 0$ leads to
  a worse performance than that achieved in Section~\ref{sec:achievable}, and so
  this cannot be the optimal parameter. We therefore have to choose
  $\e(\snr)$ such that $l_1(\snr,\e) = l_2(\snr, \e)$.  Inserting the
  definitions of $l_1$ and $l_2$ and rearranging the terms yields
  \begin{equation*}
    \snr^{-(n-1) + (n-3/2)\e} = \exp\{-\snr^\e/k\},
  \end{equation*}
  which is of the form~\eqref{eq:epsequation} with $a = -(n-1)$ and $b = n-3/2$.
  By Lemma~\ref{lem:epssolution}, for $\snr > 1$,
  \begin{equation*}
    \snr^\e = (n-3/2)k W(\snr^{\frac{2(n-1)}{2n-3}} / ((n-3/2)k)).
  \end{equation*}
  We now use the fact that $W(x)/\log x$ converges to~$1$ for $x \ra \infty$;
  this can be shown using L'H\^opital's rule and because the derivative of
  $W(x)$ is $W(x)/[x(1 + W(x))]$ (cf.~\cite{CorlessGHJK1996}).

  For sufficiently large $\snr$, therefore, there exists a constant $c > 0$ such
  that
  \begin{equation*}
    \snr^\e \ge c(n-3/2)k \left[ \frac{2(n-1)}{2n-3}\log\snr - \log((n-3/2)k)
    \right],
  \end{equation*}
  and so $\snr^\e \in \Omega(\log\snr)$. Plugging this into the bound of
  Lemma~\ref{lem:lowerbound1} we finally obtain\footnote{If $a(x) \in
  \Omega(f(x))$ and $b(x) \in \Omega(g(x))$, then $a(x)b(x)^m \in
  \Omega(f(x)g(x)^m)$.}
  \begin{equation*}
    \mse \in \Omega(\snr^{-n}(\log\snr)^{n-1}),
  \end{equation*}
  and no choice of $\e(\snr)$ can improve this bound.
\end{IEEEproof}

\section{Conclusions}

We have analyzed the source-channel coding strategy of repeatedly quantizing an
analog source and transmitting the quantizer outputs and the remaining
quantization error uncoded across $n$ Gaussian channels. We have shown that if
the quantization resolution of the encoder is chosen optimally and if the
optimal decoder is used then the mean squared error scales at best as
$\snr^{-n}(\log\snr)^{n-1}$. Furthermore, as our previous paper showed, a simple
suboptimal decoder is sufficient to achieve this scaling, so the bound is tight.

The question whether any minimal delay scheme can asymptotically perform better
than $\snr^{-n}(\log\snr)^{n-1}$ is still open at this time.

\appendices

\section{Asymptotic Notation}
\label{app:Onotation}

The ``$O$'' and ``$\Omega$'' asymptotic notation used at various points in the
paper is defined as follows.  Let $f(x)$ and $g(x)$ be two functions defined
on~$\R$. We write
\begin{equation*}
  f(x) \in O(g(x))
\end{equation*}
if and only if there exists an $x_0$ and a constant~$c$ such that
\begin{equation*}
  f(x) \le c g(x)
\end{equation*}
for all $x > x_0$. 

Similarly, we write $f(x) \in \Omega(x)$ if $\le$ is replaced by $\ge$ in the
above definition.

\section{Proof of Ziv's Lower Bound (Lemma~\ref{lem:zivbound})}
\label{app:zivboundproof}

If we condition the mean squared error on~$S$ and use the assumption on~$p_S$
we obtain
\begin{equation*}
  \mse \ge \pmin \int_A^B \msecond ds.
\end{equation*}
For $\Delta \in [0, B-A]$ we can further bound this in two ways:
\begin{align*}
  \mse &\ge \pmin \intabd \msecond ds \\
  \mse &\ge \pmin \int_{A+\Delta}^B \msecond ds \\
  &= \pmin \intabd \msecondd ds.
\end{align*}
Averaging the two lower bounds yields
\begin{multline}
  \label{eq:avglbd}
  \mse \ge \frac{\pmin}{2} \intabd \bigg( \msecond + \\
  \msecondd \bigg) ds,
\end{multline}
and applying Markov's inequality to the expectation terms leads to
\begin{equation}
  \label{eq:markov1}
  \msecond \ge \dhsq \Pr[|\Sh - S| \ge \Delta/2 \mid s]
\end{equation}
and
\begin{multline}
  \label{eq:markov2}
  \msecondd \\
  \ge \dhsq \Pr[|\Sh - S - \Delta| \ge \Delta/2 \mid s+\Delta].
\end{multline}

{\parfillskip=0pt
Now suppose that we use the communication system in question for binary
signaling. We want to send either $s$ or $s+\Delta$; at the decoder we use the
estimate~$\Sh$ to decide for~$s$ or $s + \Delta$ depending on which one $\Sh$ is
closer to. When $s$ is sent, the decoder makes an error only if $|\Sh - s| \ge
\Delta/2$; when $s + \Delta$ is sent, it makes an error only
if $|\Sh - s - \Delta| \ge \Delta/2$. The conditional error probabilities
therefore satisfy $\Pr[\text{error} | s] \le \Pr[|\Sh - S| \ge \Delta/2 \mid s]$
and
\par}

\goodbreak
\noindent
$\Pr[\text{error} | s + \Delta] \le \Pr[|\Sh - S - \Delta| \ge \Delta/2 \mid s +
\Delta]$. Applying this to~\eqref{eq:markov1} and~\eqref{eq:markov2} and
inserting the result in~\eqref{eq:avglbd} yields
\begin{equation}
  \label{eq:zivalmostproved}
  \mse \ge \pmin \dhsq \intabd \pe(s, \Delta) ds,
\end{equation}
where $\pe(s, \Delta) = \left(\Pr[\text{error}|s] + \Pr[\text{error}|s +
\Delta] \right)/2$ is the average error probability.

If $s$ and $s+\Delta$ are picked with equal probability and transmitted across
$n$~parallel Gaussian channels as $\X(s)$ and $\X(s+\Delta)$, and if $d(s,
\Delta) = \| \X(s) - \X(s + \Delta)\|$, then the error probability of the MAP
decoder is $Q(d(s,\Delta) / 2 \sz)$, a standard result of communication theory
(see \eg~\cite[Section~4.5]{WozencraftJ1965}). Because the MAP decoder minimizes
the error probability, $Q(d(s,\Delta)/2\sz) \le \pe(s,\Delta)$, which, when
inserted into~\eqref{eq:zivalmostproved}, completes the proof. \hfill\IEEEQED

\bibliography{mkbiblio}
\end{document}